# *Topological Valley Photonic Waveguides: Scattering matrix evaluation for linear computing*

*Christian Johnson-Richards[1,] Alex Yakovlev [2] and Victor Pacheco-Peña [1,*]*

[1] *School of Mathematics, Statistics and Physics, Newcastle University, Newcastle Upon Tyne, NE1 7RU, United Kingdom*
[2] *School of Engineering, Newcastle University, Newcastle Upon Tyne, NE1 7RU, United Kingdom*

*\*email: victor.pacheco-pena@newcastle.ac.uk*

**Abstract**

Topological boundary modes utilizing valley mode waveguides have opened opportunities in, for instance, the design of high transmission waveguides with tolerance to geometrical defects and sharp bends. Applications of these waveguides include linear computational processes and the emulation of logic gates using linear structures, among other scenarios. Here we present the design of a 6-port junction that exhibits equal power splitting to three other ports when excited at single port with no reflections. In studying this structure, a scattering matrix is extracted at telecom wavelengths (around 1550 nm). The linearity of the system along with the scattering matrix are exploited to produce linear operations such as routing of information considering two incident signals or multiple signals applied from different ports. Our work may be exploited to analytically design larger networks without the need of computationally expensive trial and error numerical methods.

# 1. Introduction

The arbitrary manipulation of electromagnetic (EM) waves is at the core of many research fields including the photonics and metamaterials communities [1], [2]. Efforts in these areas have offered new and alternative ways to manipulate wave-matter interactions both in space (via spatial inhomogeneities along the path where a wave travels) [3], [4], [5], [6], [7] and recently in time [8], [9], [10] and/or space-time [9], [11], [12], [13]. These efforts have enabled further opportunities for the proposal of new or innovative improvements of already existing applications, including antennas [14], [15], [16], sensing [4], lensing [17], [18], holography [19], unidirectional transport [20], topological photonics [21] and computing with waves [22], [23], [24].

The field of topological photonics finds its roots in a seminal work where it was shown how quantum-spin hall effects [25], [26], [27], [28], [29], [30] can be translated to the field of photonics in the form of augmented photonic crystals (PhCs), named Photonic Topological Insulators (PTIs) [31], [32]. PTIs provide resistance to the scattering and back-reflection from lattice defects and sharp bends of a signal travelling along a waveguide [21]. Such "topological transport" has attracted the attention of the scientific community due to the potential use in the design of low loss compact photonic circuits. This is due to the increased transmission of the signal traveling through waveguides with defects and/or sharp bends. Applications exploiting PTIs include routing [33], [34], [35], power splitting using four port junctions [36], [37], [38], [39], [40], linear Boolean operations such as XOR/OR [41], quantum computing [42], and lasing [43], [44], among others. To design a PTI, one can choose a PhC lattice with degenerate wavevector and induce a photonic bandgap. To induce a band gap, one can break time reversal symmetry [45], [46], [47], [48]. Chern PTIs are one example of this and have been shown to display a single unidirectional wavevector for a propagating mode along the boundary with other PhCs or air [49]. Alternatively, one can break spatial inversion symmetry [47] by using Valley Photonic Crystals (VPCs) [21], [50], [51], [52], [53], [54], [55], [56]. VPC-VPC waveguides can be formed at the boundary between two VPCs with opposing pseudospin/valley-Chern number and support a single forward and backward propagating wavevector [57], [58]. VPC-VPC waveguides have recently been applied in hybrid topological photonic crystals [59], slow light [60], plasmonics [61] and non-Hermitian phenomena [62]. The present manuscript is focused on a dielectric VPC for computing.

Computing with EM waves has shown to be a promising candidate to enable high-speed calculations (at the speed of light within the involved materials). By exploiting waveguides working as ideal transmission lines, applications such as routing, analogue, quasi-digital computing operations and partial differential equation solving have been recently proposed [23], [63], [64], [65], [66]. Inverse designed metamaterial-based photonic devices have also been demonstrated and applied to computing. For instance the calculation of mathematical operations such as differentiation and integration directly on the temporal envelope of the signal [8], [22], [67], [68]. Further scenarios of wave-based computing include sub-wavelength dielectric scattering structures that solve equations [69], [70], an integrated photonic hardware accelerator [71], signal processing [72], and elements that bridge Kirchhoff's law from circuit theory to photonics [73], to name a few.

Inspired by the opportunities of EM computing and the importance of topological photonic structures, here we present our efforts in designing a 6-port junction constructed with VPC-VPC waveguides working at the telecom frequencies of 193.4145 THz ($\approx$ 1550nm). The structure was designed using two types of VPC (namely type I and type II), with ports positioned at angles of $\pi/3$ radian intervals. In this way each port has the same VPC-VPC interface (as it will be shown below). The proposed 6-port VPC junction was numerically evaluated demonstrating that, when one input is used to apply the incident signal, equal power splitting is observed towards 3 output ports with zero power reflected towards the input port. The full scattering matrix of the structure is then calculated within the spectral range of ≈ 187.1 to ≈ 199.9 THz (within the band-gap spectral window). The extracted scattering matrix proves critical when designing photonic circuits and linear operations as it avoids trial and error numerical design that is computationally costly. To demonstrate the potential of the proposed 6-port VPC junction and the extracted scattering matrix analytically representing it, two examples of linear computing devices are presented: I) a device where two simultaneous inputs are redirected towards two predefined output ports by exploiting the phase difference between the two incident signals and II) a *wave director* where all the signals applied to the input ports are directed towards a single port without reflection.

## 2. Valley Photonic Crystal Design and Analysis

In this section, the design and topological properties of the chosen VPC is presented. As a first step, its performance is evaluated by studying the influence of geometrical parameters that affect the bulk band structure. From this, four different combinations of parameters (namely unit cell periodicity and radius of the holes) are selected in order to produce a bandgap centered on the desired operating frequency, $f_{\text{design}} \approx 193.4$ THz (telecom wavelength of 1550 nm). The next step involves the study of the VPC-VPC waveguide super-cell band structure where it will be shown how there exists a uniquely guided wavevector and how the surrounding band features may affect the topological properties. From this, a unit cell periodicity and radius of the holes are selected to construct the 6-port junction.

### 2.1 Bulk Band Structure Study

The proposed VPC is designed using a hexagonal lattice in order to support degenerate wavevectors at the $K$ and $K'$ points [47]. A schematic diagram of four-unit cells in the direct space and one Brillouin zone in the reciprocal space is presented in the left and right panels from Figure 1a, respectively. The lattice vectors are labeled as $a_1$ and $a_2$, and the reciprocal lattice vectors as $b_1$ and $b_2$. The unit cell is made of silicon as the background material (permittivity of $\varepsilon_{Si} = 12$ [74]) and two air holes ($\varepsilon_{Air} = 1$) with radius $r_1$ and $r_2$ centered at positions $\mathbf{h_1}$ and $\mathbf{h_2}$, respectively (see the caption of Figure 1 for the specific values). A frequency independent permittivity for silicon is used due to its small variation at telecoms frequencies, in this way reducing the complexity of the model [33], [41], [74]. As shown in the schematic of the reciprocal space from Figure 1a, the high symmetry wavevectors are labeled as $\mathbf{K,K',M}$ and $\mathbf{\Gamma}$. With this configuration, the numerical results of the first three transverse electric (TE) bands (in-plane $E$-field [75]) are presented in Figure 1b (see methods section for details on the simulation setup). Here, two cases are considered: first, a case when spatial inversion symmetry is maintained by choosing $r_1 = r_2$ (blue dashed lines with the first two bands in light blue and third band in dark blue). In this case a Dirac-cone where two dispersion curves meet is present at the $\mathbf{K}$ and $\mathbf{K'}$ lattice vectors (i.e., no bandgap). The second scenario represents the case when spatial inversion symmetry is broken by setting $r_2 \neq r_1$. To do this, we define the hole size ratio $\Omega = r_2/r_1 = 1/2$, noting that varying $\Omega$ here means fixing $r_1$ and changing $r_2$. This is presented as solid red lines in Figure 1b, with the first two bands in light red and the third in dark red. From these results, it can be seen that a bandgap is induced at the $\mathbf{K}$ and $\mathbf{K'}$ point (with its extent highlighted with a light red box) covering the frequencies of $\approx 187.5$ THz to $\approx 206.0$ THz ($\approx 1455.3$ nm to $\approx 1598.9$ nm). For completeness, the $H_z$ field at the lower $\mathbf{K(K')}$ and upper $\mathbf{K(K')}$ wavevectors (labelled i(iii) and ii(iv) respectively in Figure 1b) were calculated and the corresponding phase distributions are shown in Figure 1c(i,iii) and Figure 1c(ii,iv) respectively. The phase vortices observed in Figure 1c are a hallmark of the orbital angular momentum that produces the non-zero valley-Chern number required for VPCs, suggesting that it is possible to observe topological transport with this VPC design [76]. Additionally, it is important to note how the phase vortices alternate direction between both the $\mathbf{K}$ ($\mathbf{K'}$) wavevectors as well as the lower and upper bands. This can be seen in Figure 1c(i,iii) and Figure 1c(ii,iv) where an anti-clockwise and clockwise rotation at the position $\mathbf{h_1}$ and $\mathbf{h_2}$ occurs, respectively, as expected [47].

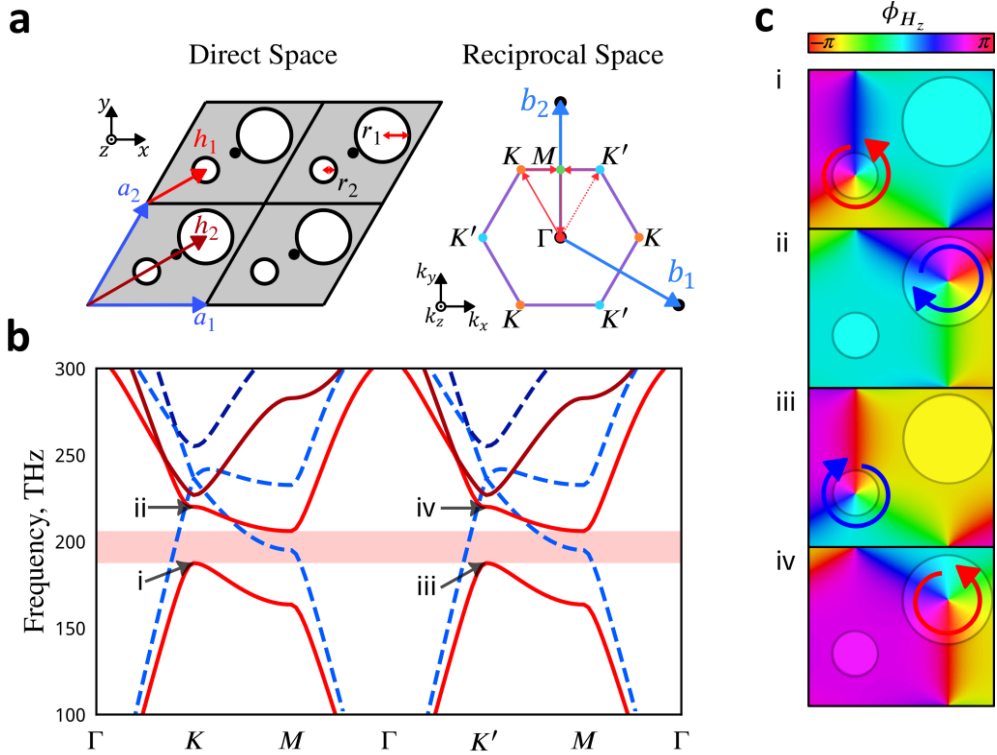

**Figure 1.** VPC bulk band study. (a) Left panel depicts a type-I VPC hexagonal unit cell in direct space, with the lattice constant $a = 0.34$ μm and lattice vectors $\boldsymbol{a_1} = (a, 0)$ and $\boldsymbol{a_2} = (a/2, a\sqrt{3}/2)$ with a hole ratio, $\Omega = 1/2$ ($r_1 = 0.24a$), each with a hole centered on the vectors $\boldsymbol{h_1} = 1/3(\boldsymbol{a_1} + \boldsymbol{a_2})$ and $\boldsymbol{h_2} = 2/3(\boldsymbol{a_1} + \boldsymbol{a_2})$. The right panel represents a schematic of the reciprocal space with reciprocal lattice vectors $\boldsymbol{b_1} = (1/a, -\sqrt{3}/3a)$ and $\boldsymbol{b_2} = (0, 2\sqrt{3}/3a)$ with high symmetry points highlighted at the corners of the Brillouin zone. (b) Numerically calculated bulk TE band structure for the first 3 bands for two cases: a non-topological PhC ($\Omega = 1.0$) is presented in blue dashed lines and a VPC ($\Omega = 0.5$) is shown as red solid lines. The band gap between the first and second band is highlighted in a red box. (c) Phase distribution $\phi_{H_z}$ across the unit cell for the first and second band at the $\boldsymbol{K}$ [labelled i and ii in panel (b)] and $\boldsymbol{K'}$ [labelled iii and iv in panel (b)] wave-vectors respectively.

The origin of topological protection in VPCs is undoubtably the induced band gap created when spatial inversion symmetry is broken [47]. This band gap can be tuned by carefully selecting the geometrical parameters such as $a$, $\Omega$ and the materials of the VPC. It can then be characterized in terms of the band gap central frequency ($f_{\text{BG}}$) and bandwidth ($\Delta f_{\text{BG}}$) [77]. In this section we evaluate $f_{\text{BG}}$ and $\Delta f_{\text{BG}}$ for a set of VPCs by varying $a$ and $\Omega$ (see Figure 2 caption for the range of values) while fixing the background and hole materials to silicon and air respectively. The results describing the change in $f_{\text{BG}}$ and $\Delta f_{\text{BG}}$ are presented in Figure 2a and Figure 2c respectively, noting that the colourmap chosen for Figure 2a is set so that $f_{\text{design}}$ is located at the centre of the scale (white color). To better study the results shown in Figure 2a,c, we extracted $f_{BG}$ and $\Delta f_{BG}$ as a function of $a$ for four values of $\Omega$, namely $\Omega_i = 0.275$, $\Omega_{ii} = 0.5$, $\Omega_{iii} = 0.634$ and $\Omega_{iv} = 0.733$ and the results are shown in Figure 2b,d. First, looking at Figure 2a,b it is clear that $f_{BG}$ is shifted when $a$ is varied for all four values of $\Omega$. For example, in Figure 2b when $a$ is varied within the full range here studied (250 to 410 nm) an overall shift of $f_{BG}$ of $\approx$ 100.5 THz is obtained at $\Omega = \Omega_i = 0.275$. Similarly, an overall shift of $f_{BG}$ of $\approx$ 110.1 THz is observed for $\Omega = \Omega_{iv} = 0.733$. Additionally, it can be seen in Figure 2a that changing $\Omega$ and keeping $a$ constant has a smaller effect on $f_{BG}$. This is expected as $a$ represents the scaling of all dimensions in the design of the VPC, whereas changing $\Omega$ only scales $r_2$.

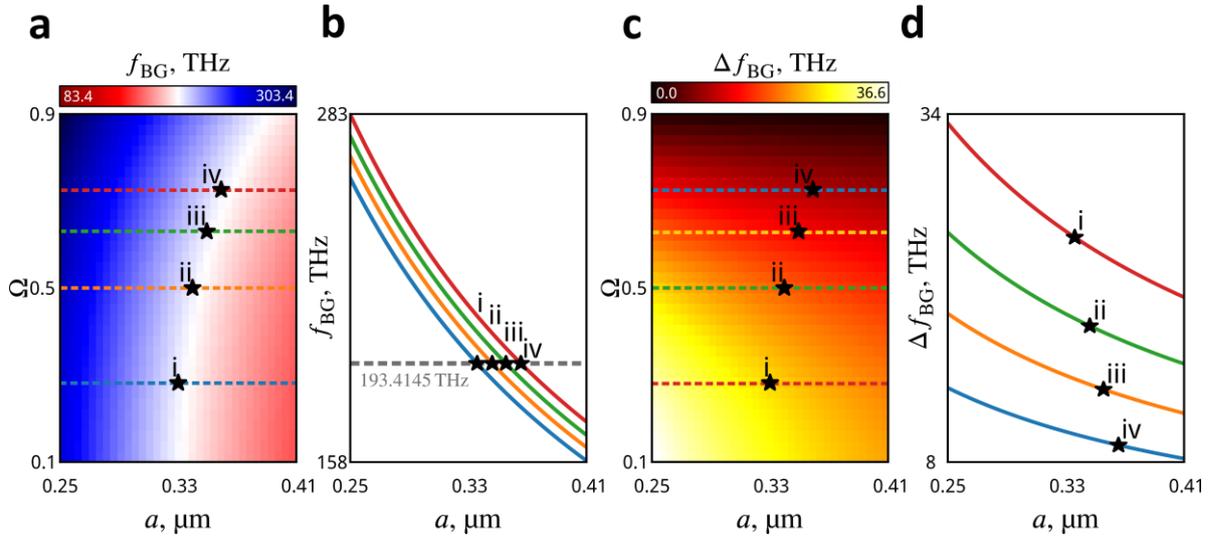

**Figure 2.** Study of lattice constants, $0.25 \leq a \leq 0.41$, and hole ratios, $0.1 \leq \Omega \leq 0.9$ effect on $f_{BG}$ and $\Delta f_{BG}$. (a) $f_{BG}$ and (c) $\Delta f_{BG}$ for the first band gap (highlighted in red in Figure 1b). (b) Line plots of horizontal slices labelled i-iv from panel a, for values of $\Omega_i = 0.275$ (blue), $\Omega_{ii} = 0.5$ (orange), $\Omega_{iii} = 0.634$ (green) and $\Omega_{iv} = 0.733$ (red). The star symbols in the figure represent $\Omega$ and $a$ combinations that result in the same $f_{BG}$, specifically $\Omega_{i-iv}$ and corresponding values of $a_i = 0.33$ μm, $a_{ii} = 0.34$ μm, $a_{iii} = 0.35$ μm and $a_{iv} = 0.36$ μm, respectively. (d) Same as panel b but for the results from panel c.

Now drawing our attention to Figure 2c,d, it is observed that increasing $a$ has the effect of decreasing $\Delta f_{BG}$ at all four values of $\Omega$ shown in Figure 2d. Quantitatively, for example, from the results shown in Figure 2c, when $a$ is varied within the same range as in Figure 2a,b, $\Delta f_{BG}$ varies $\approx$ 12.9 THz, $\approx$ 9.6 THz, $\approx$ 7.2 THz and $\approx$ 5.24 THz for each $\Omega_{i-iv}$, respectively. These findings are consistent with [35], [41], [61], showing that modifying $a$ can be used to shift the central frequency of the bandgap, while $\Omega$, which is related to the degree of which inversion symmetry is broken, can be used to modify $\Delta f_{BG}$ [77]. Interestingly, one can select different combinations of $a$ and $\Omega$ that produce a bandgap at the same center frequency (as shown by the white region in Figure 2a). In this case, for the values of $\Omega_{i\text{-}iv}$ as described above, the values of $a$ that produce the same $f_{BG}$ are represented as star symbols in Figure 2 (see specific values in the caption of this figure). These values will be used in the next section to evaluate the band structure of a VPC-VPC waveguide.

### 2.2 Waveguide Supercell Band Structure Study

Here we explore VPC-VPC waveguides designed using the VPCs from the previous section. We name VPC-1 and VPC-2 each consisting of the VPCs forming the waveguides with VPC-2 being the mirror symmetric version of VPC-1 in the *x*-axis. As mentioned in the introduction, this is necessary to produce a change in the valley-Chern number across the boundary between the two VPCs, in this way inducing an edge state that supports topological transport [47] (localized at the bandgap of the constituent VPCs [52]). As VPC-2 is the mirror-symmetric version of VPC-1, there are two possible orientations of the resulting VPC-VPC interface of the waveguide: type I, where the two larger holes of each VPC are adjacent at the interface of the waveguide and type II, where the two smaller holes of each VPC are adjacent across the interface. The type I interface was selected after initial investigations of the 6-port junction revealed that a waveguide designed with type II orientation allowed for multiple valid wavevectors (see supplementary information section 1) at its corresponding equal power splitting frequency, whereas type I supports a unique wavevector, as it will be discussed below.

To begin with, let us study the VPC-VPC waveguide band structures using a supercell [75] with dimensions considering the four cases (i-iv) as discussed in the previous section. i.e., each supercell is designed with the corresponding pairs of lattice parameters $a_{\text{i-iv}}$ and $\Omega_{\text{i-iv}}$, and with dimensions of exactly $a$ along the horizontal $x$-axis and $43a$ along the vertical $y$-axis (this latter dimension was chosen so the evanescent fields along the vertical $y$-axis had sufficiently decayed before encountering the upper and lower boundaries). A schematic of a supercell is presented in Figure 3b(i) (cropped to $20a$ along the vertical $y$-axis to better appreciate its features) where the black/white regions correspond to silicon/air materials, respectively. For more information on the model see the methods section. With this configuration, the TE band structures of the periodic supercells are then numerically calculated for horizontal wavevectors $-0.5 \leq k_x \leq 0.5$ and solutions in the spectral window $155 \leq f \leq 225$ THz for all four cases. The numerical results are shown in Figure 3a(i-iv). The edge state that supports topological transport is represented in each panel as dark blue lines and regions of interest have been highlighted with the following colors: grey, purple, pink, green and yellow, corresponding to the upper/lower bulk wavevectors, a band gap between the upper bulk and the topological wavevectors, multiple valid topological wavevectors, unique topological wavevectors and radiative wavevectors situated above the light line, respectively.

As observed in Figure 3a(i-iv), as the lattice constants are varied from case i to case iv, the edge states (blue line) are spectrally stretched, corresponding to a spectral width (calculated considering the highest and lowest frequency value where the topological edge states, blue lines, exist) of $\approx$ 26.78 THz, $\approx$ 30.44 THz, $\approx$ 33.08 THz, $\approx$ 35.08 THz for each case, respectively. This is due to the edge states becoming less localized as the bandgap of the constituent VPCs is reduced (i.e., by increasing the hole ratio $\Omega$, as seen in Figure 2d) [52], [75]. Moreover, because of the reduced VPC bulk bandgap, the overall spectral bandwidth between the upper and lower bulk wavevectors also decreases (which are slightly perturbed from the VPC bulk bandgaps due to the discontinuity in periodicity at the interface [34]) with values of $\approx$ 26.47 THz, $\approx$ 19.93 THz, $\approx$ 14.95 THz, $\approx$ 11.02 THz for each case, respectively. Interestingly, the bandgap (purple region in Figure 3a) only exists for cases i and ii, with spectral widths of $\approx$ 6.60 THz and $\approx$ 2.15 THz respectively, suggesting a threshold of $\Omega$ for this bandgap to exist, as in [61]. Regarding the band where multiple edge state wavevectors exist (pink region) their spectral width is of $\approx$ 4.11 THz, $\approx$ 4.32 THz, $\approx$ 3.85 THz and $\approx$ 2.30 THz respectively, noting that this band is not used here as we are interested in edge states with unique wavevectors. The band with unique edge state wavevectors (green region) is observed to vary in spectral bandwidth for each case, with values $\approx$ 10.68 THz, $\approx$ 12.81 THz, $\approx$ 11.06 THz and $\approx$ 8.72 THz respectively. Finally, the radiative bands (yellow region) reduce for each case, existing only in cases i-ii (similar to the threshold observed for the purple region, as mentioned above) with bandwidths of $\approx$ 5.09 THz and $\approx$ 0.64 THz respectively.

As mentioned above, the bandwidth of the bulk VPC bandgap can be related to the localization/topological protection of the edge states of a VPC-VPC waveguide [34], [42], [47], [52]. Hence, if case i is selected (with the largest bulk bandgap) one should expect to obtain more topological protection of the edge states when compared to scenarios with smaller bulk bandgaps. On the other hand, the bandwidth of the unique edge state wavevectors is largest for case ii and when looking at how the central frequency of the uniquely guiding band varies from $f_{\text{design}}$ for the VPC-VPC waveguide, for the cases shown in Figure 3, this variation is of $\approx$ 0.26 THz, $\approx$ 0.06 THz, $\approx$ 0.90 THz and $\approx$ 1.58 THz, respectively. Based on these considerations, case ii ($a_{\text{ii}} = 340$nm and $\Omega_{ii} = 0.5$) was chosen as the best tradeoff between the topological protection provided by large bulk bandgaps and the deviation of the central frequency of the green region band (Figure 3) from $f_{\text{design}}$. With the selected lattice parameters, for completeness, the magnetic field magnitude and phase distribution of an example VPC-VPC edge state is presented in Figure 3b(ii-iii) respectively, demonstrating how the edge state mode mostly travels along the interface between the two VPCs, as expected. The design of the 6-port structure using such edge state will be discussed in the next section.

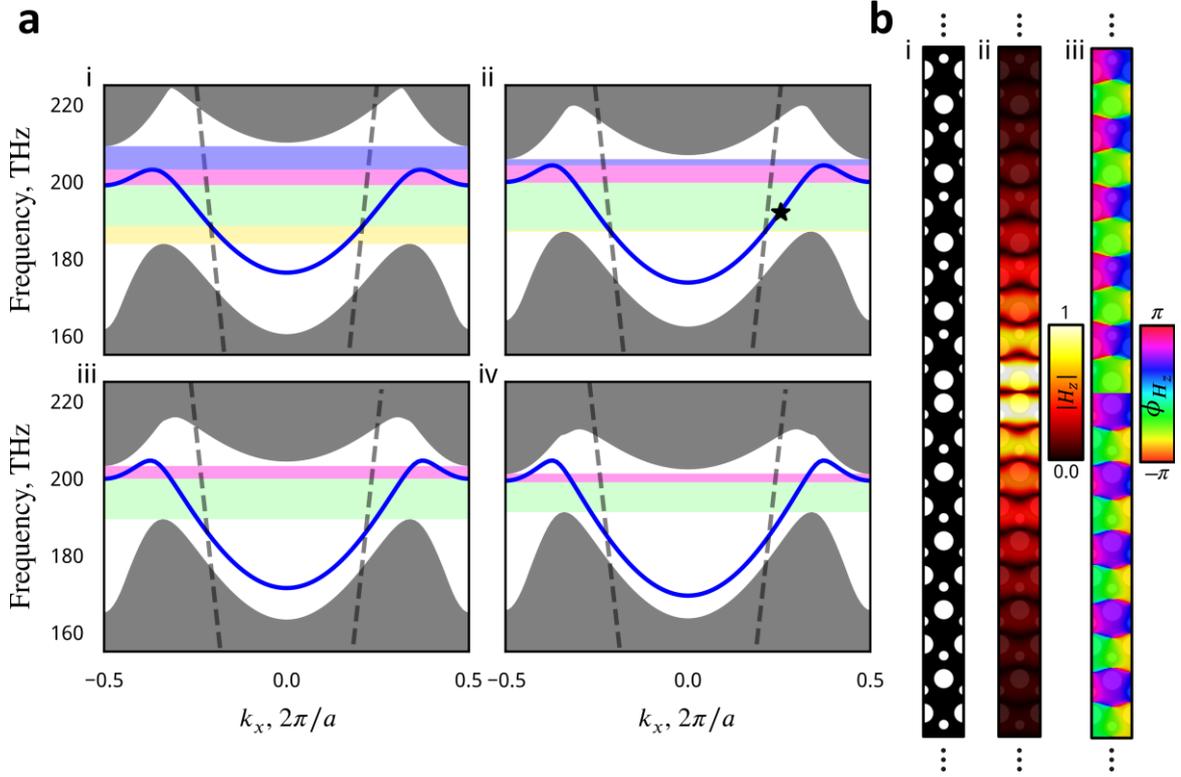

**Figure 3.** VPC-VPC type I study. (a) band structures of four supercells designed with pairs of lattice parameters $a_{\text{i-iv}}$ and $\Omega_{\text{i-iv}}$, panels i-iv. The grey, purple, pink, green and yellow regions correspond to the upper and lower bulk wavevectors, a band gap, multiple wavevectors, unique wavevectors and radiative wavevectors above the light line (dashed grey lines) respectively. (b) i) schematic of the VPC-VPC waveguide of horizontal length $a$ and vertical length $43a$ (cropped to vertical height of $20a$). The white region corresponds to $\varepsilon_{\text{Air}}$ and the black region to $\varepsilon_{\text{Si}}$. ii-iii) magnitude, $|H_z|$, and phase distribution, $\phi_{H_z}$, of the magnetic field, across the supercell for the "star" marked wavevector in panel (a)ii at a frequency of $\approx 192$ THz and $k_x \approx 0.25$.

## 3. 6-Port Equal Power Splitter

In this section we design and study a 6-port VPC junction using the previously selected VPC lattice parameters. It will be shown that the junction (central region of the structure) can support equal power splitting of a signal incident at, for instance, port 1 to ports 2, 4 and 6. The scattering matrix is also extracted in order to use it for linear operations (as it will be discussed later) [34]. To design the 6-port junction, six type-I VPC-VPC waveguides are rotated by π/3 radians clockwise around the junction center corresponding to ports 1-6. A schematic of the junction center is presented in Figure 4a(ii) outlining each of the hexagonal unit cells and holes. VPC-1 and VPC-2 are highlighted in grey and orange color, respectively (noting that the materials are still silicon and air as in the previous sections).

As an initial step, we can heuristically interpret the transport of incident signals of this junction by considering which VPC-VPC edge state is being excited [42]. Consider that there are two types of edge states, a $K'$ and $K$-type, which correspond to the forward and backward propagating states in Figure 3a. The propagation of each edge state is supported on either side by out of plane magnetic field phase vortices in the bulk VPC [78] (similar to those presented in Figure 1c) that spin in opposing directions in VPC-1 and VPC-2 as well as each $K$-type. As a result, the coupling between the $K$ and $K'$ states is incompatible [42], [47]. To visualize this with the 6-port structure, consider an incident signal applied from port 1. At this port, the waveguide consists of two VPCs: VPC-1 (top) and VPC-2 (bottom). When the incident signal reaches the junction, the waveguide towards port 2, 4 and 6 will have the same VPC-1/VPC-2 direction as the waveguide connected to port 1 ($K'$ type). On the other hand, the waveguides toward ports 3, 5 and towards port 1 (for this latter waveguide when moving from right to left) have the opposite orientation ($K$-type). Because of this, it is expected that an input signal applied from port 1 will scatter at the junction, producing signals traveling only towards ports 2, 4 and 6, with forbidden propagation towards the rest of the ports (ports 1, 3 and 5). A schematic of the expected coupling is presented in Figure 4a(i), where ports 1 to 6 are labelled $P_{1-6}$, the red arrows represent the allowed coupling of a signal incident at $P_1$ and black crosses representing forbidden coupling.

With this configuration, the numerical results of the out of plane magnetic field, $H_z$, distribution for an incident signal applied at $P_1$ with an incident frequency of ≈ 192.08 THz is shown in Figure 4b (see details in the methods section; as it will be shown below, this frequency corresponds to the frequency where equal power splitting occurs). For this case, we consider the 6-port junction to have a finite size defined by a hexagonal boundary (green hexagon) of side length $d_{\text{hex}} = 19.5a$ surrounded by a square buffer region of side length $d_{\text{sim}} = 63a$. The value of $d_{\text{hex}}$ is chosen so that the retrieved magnitude and phase (discussed below) only consider scattered light at the junction that has coupled into each respective waveguide (a study of the $d_{\text{hex}}$ dependence on the retrieved scattering parameters is presented in the supplementary information section 3). The value of $d_{sim}$ is chosen to allow space for a point dipole source placed at the edge of the buffer region (see methods for the exact positioning) to couple into the edge state before entering the junction (green hexagon in Figure 4a-b). As observed, the incident signal propagates towards ports $P_{2,4,6}$ with no observable transmission to ports $P_{3,5}$ or reflection back towards $P_1$, as expected.

To further characterize the performance of the designed 6-port structure, the scattering parameter $|S|^2$, and phase, $\phi$, for all the ports when considering an incident signal from $P_1$ are shown in Figure 4c (top and bottom panels, respectively). Results are displayed for 200 equally spaced frequencies in the spectral range of 185 THz to 201 THz (see methods section for details of the extraction of the scattering parameters). Note that the phase for $P_{1,3,5}$ is omitted as there is no signal transmitted towards these ports. The points (crosses) in the scattering parameter (phase) dataset represent numerical simulations with the solid lines representing a linear fit. Moreover, the wavevector bands discussed in section 2.2 are highlighted in both panels with the same color scheme as in Figure 3. Note that results for frequencies outside the single wavevector band (green) are not considered for the linear fit. As observed, within the green region (unique wavevector band) there is approximately zero transmission to $P_{3,5}$ or reflection back to $P_1$ (from now on, these will be considered as zero). Within the radiative (yellow), bulk (grey) and multiple wavevector (pink) bands there are some measurable reflections to $P_1$, as expected [34]. From these results, the junction exhibits equal power splitting, interpolated from the linear fit in Figure 4c, at a frequency of $f_s \approx 192.08$ THz, with values of $|S|^2$ of $\approx 0.33$ for ports $P_{2,4,6}$. For completeness the retrieved $\phi$ values at the frequency $f_s$ are $\approx -1.80$, $\approx 0.29$ and $\approx -1.80$ (all radians) for $P_{2,4,6}$ respectively. As discussed in the previous sections, the dimensions of the VPCs were tuned to place the center of the unique wavevector band at a frequency of $f_{design}$. Interestingly, the equal power splitting frequency ($f_s$) is near $f_{design}$. A potential reason of this performance may be that since $f_{design}$ is at the center of the unique wavevector band, the excited edge state has the most topological protection when compared with other edge states that still fall within the green region from Figure 3a(ii) but at different frequencies. This is because such edge states are much closer to the radiative or nonunique wavevector bands. Such protection of the edge state at the frequency $f_{\text{design}}$ may translate into an ideal behavior of the 6-port structure, enabling equal power splitting where only propagation towards ports $P_{2,4,6}$ is allowed. The small deviation between $f_s$ and $f_{\text{design}}$ may be due to the junction (central region of the structure) which breaks, even slightly, the periodicity of the waveguides. This phenomenon, however, would require further study and it is outside the scope of the current manuscript.

Now, moving away from $f_s$, when working at lower frequencies, propagation towards $P_4$ increases. This may be because the operating frequency becomes closer to the radiative (yellow) region, affecting the coupling of the edge state for the waveguides connected to $P_{2,6}$. On the other hand, when the incident frequency is increased, propagation towards $P_4$ decreases. This is because the central region (junction) of the structure becomes a more relevant defect (smaller incident wavelength). Hence, the signal is scattered and more energy is directed towards $P_2$ and $P_6$. (For completeness the magnetic field distribution at different frequencies is shown in the supplementary information document section 2). An additional investigation where holes are removed from the center of the junction, which effectively increases the size of the junction defect, is presented in the supplementary information section 4). To better observe the results shown in Figure 4c, the retrieved scattering parameters were used to produce a full $6 \times 6$ frequency dependent scattering matrix as the 6-port structure is both rotationally symmetric and reciprocal [79]. A representation of the $6 \times 6$ scattering matrix for the $|S|^2$ and $\phi$ values is presented in Figure 4d, where the upper row represents $|S|^2$ and the lower row represents the $\phi$ at three frequencies labelled i-iii (vertical dashed lines in Figure 4c). As observed, at the frequency of equal power splitting (middle column from Figure 4d), transmission towards ports $P_{2,4,6}$ is obtained when the incident signal is applied at $P_1$. The same occurs when the signal is applied from $P_2$, where transmission is obtained towards $P_{1,3,5}$. Due to this *rotation* of the scattering parameters, the 6 × 6 map at this frequency resembles a checkerboard, which is not what occurs at other frequencies, as shown in the same Figure 4d. With the scattering parameters obtained, we can define linear expressions (valid in the unique topological wavevector band) that fit the results from Figure 4c (units of frequency in THz), as follows (note that, as mentioned above, the signals traveling towards $P_{1,3,5}$ are negligible):

$$|S|^2_{P_2} = (0.01404)f - (2.36372) \tag{1a}$$

$$|S|^2_{P_4} = -(0.03226)f - (6.52928) \tag{1b}$$

$$|S|^2_{P_6} = (0.01404)f - (2.36372) \tag{1c}$$

$$\phi_{P_2} = (1.76089)f - (327.457) \tag{1d}$$

$$\phi_{P_4} = (1.83243)f - (345.396) \tag{1e}$$

$$\phi_{P_6} = (1.76108)f - (327.493) \tag{1f}$$

With the extracted scattering matrix, we can now analytically model the splitting at a 6-port topological junction with arbitrary inputs. In the next section we leverage the extracted scattering matrix to investigate the 6-port structure when excited with two inputs as well as design two arbitrary input devices. It is important to note that the scattering values shown in Figure 4 correspond to the case when the size of the whole structure is $d_{\text{hex}} = 19.5a$. For completeness, the scattering parameters for a 6-port junction of size smaller and larger than this $d_{\text{hex}}$ are provided in the supplementary materials section 3.

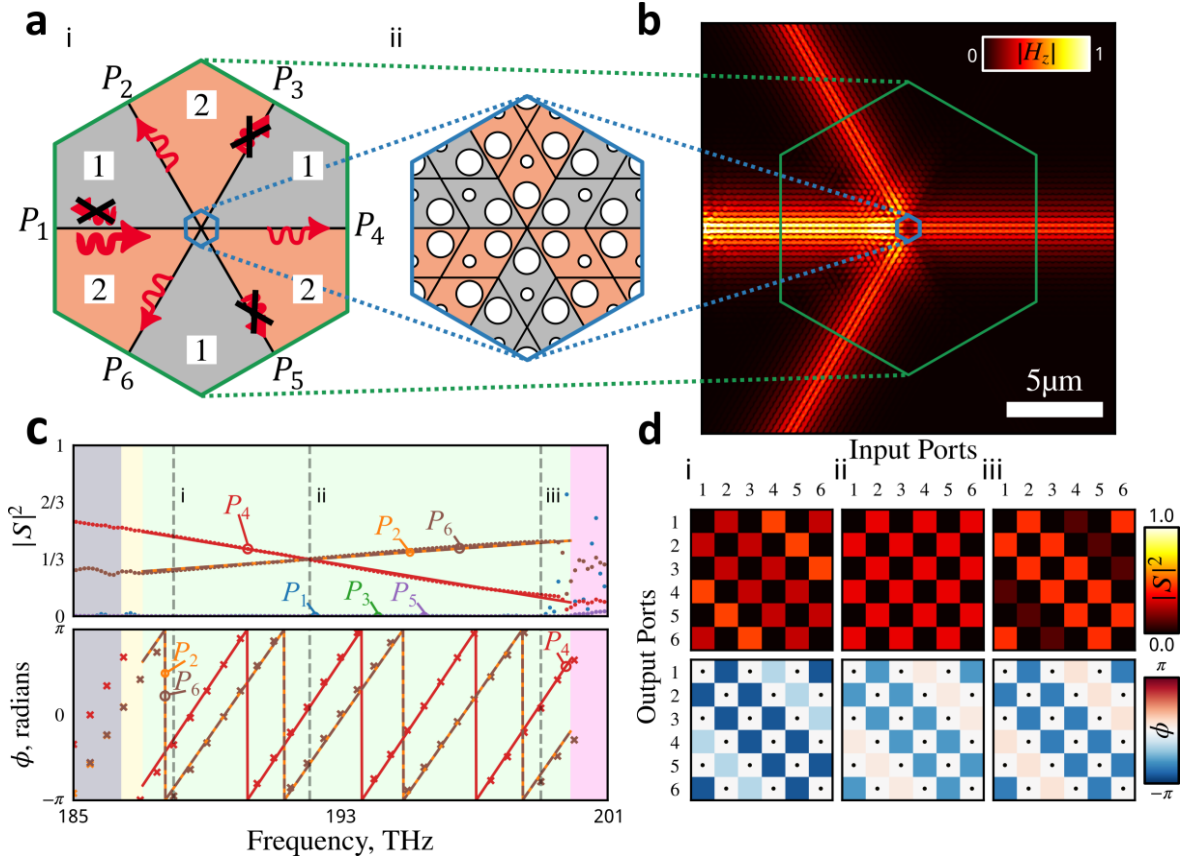

**Figure 4.** 6-Port VPC-VPC junction study. (a) schematic representation of the designed structure using VPC-1 (grey) and VPC-2 (orange) labelled as 1 and 2, respectively. Panel i the red arrows represent the splitting of an input signal applied from $P_1$ to $P_{2,4,6}$. Panel ii shows a zoom-in schematic of the central region displaying the outline of each unit cell. (b) Numerical result of the magnetic field distribution, $|H_z|$, at an input frequency of $f_s \approx 192.08$ THz with an input signal at $P_1$. The green hexagon denotes the size of the device shown in (a)i and the blue hexagon outlines the central region shown in (a)ii. (c) Retrieved scattering parameters for a single input at $P_1$ with the supercell bands overlayed as colored regions (see Figure 3 caption for details). The vertical dashed lines marked as i-iii correspond to the frequencies 188 THz, $f_s$ and 199 THz respectively. (d) Checkerboard representation of the retrieved scattering matrix of the 6-Port junction at the frequencies marked i-iii in panel (c). The dots in the phase checkerboards represent ignored phases (as there is no transmission towards those ports).

## 4. Applications

In the previous section we extracted a frequency dependent scattering matrix, referred to as $M_S$ from now on, that represents the splitting of signals at a 6-port topological junction of size $d_{\text{hex}} = 19.5a$. In this section we make use of $M_S$ to tailor the output signals using carefully engineered inputs. As a first step we define an input vector, $X = [x_{P_1}, \ldots, x_{P_6}]$ where $x_{P_m}$ represents the signal incident at $P_m$ in phasor form, of length 6. i.e., this vector corresponds to the amplitude and phase of a signal applied at each port, $P_{1-6}$, respectively [63], [80]. For example, if the incident signals are applied using two inputs (say $P_2$ and $P_6$) with a phase of $\phi_{P_2}$ and $\phi_{P_6}$, respectively, the input vector is $X = [0, 1e^{i\phi_{P_2}}, 0, 0, 0, 1e^{i\phi_{P_6}}]$. As a result, an output vector of signals, $Y = [y_{P_1}, \ldots, y_{P_6}]$ where $y_{P_m}$ represents a phasor of the output signals at $P_m$, of the 6-port junction can be simply calculated as $Y = M_S X$ [63], [65], [79]. With this in mind, by fixing $\phi_{P_2} = 0$ radians and changing $\phi_{P_6}$ for a two-input scenario, the output magnitude for each port, $|y_{P_m}|^2$ (where $m = 1, 2 \ldots, 6$), is evaluated using this linear approach is shown in Figure 5a for frequencies $187 \leq f \leq 199$ THz (within the unique wavevector band of the VPC-VPC wavevector). In this figure, each panel represents $|y_{P_m}|^2$ at each port as a function of frequency (vertical axis) and phase difference $\phi_{P_6} - \phi_{P_2}$ (horizontal axis). From these results, no signals propagate towards $P_{2,4,6}$. This is expected as if one considers the input signals from $P_2$ and $P_6$ separately, they will only generate output signals towards $P_{3,5,1}$ and $P_{1,3,5}$ respectively. At $f = f_s \approx 192.08$ THz and $\phi_{P_6} = 0$ radians, the signals are directed towards $P_{1,3,5}$ with values of $|y_{P_{1,3,5}}|^2 \approx 1.33, \approx 0.34$ and $\approx 0.34$, respectively. Interestingly at $f = f_s$ and $\phi_{P_6} - \phi_{P_2} = \pi$ radians, the signal is split equally between $P_3$ and $P_5$ with corresponding values of $|y_{P_{3,5}}|^2 \approx 0.9937$ for each port. As observed, the value is near but not exactly 1, which could be attributed to, for instance, a numerical error due to finite meshing and/or small reflections still present from the absorbing boundaries that are being coupled into the forbidden edge states, leading to the measurement of a lossy scattering matrix. Alternatively, and more fundamentally, this small difference could be attributed to the junction itself not being an idealized structure. i.e., as it has been shown in transmission lines at microwave frequencies, perfect splitting can be obtained when using ideal transmission lines where, for instance, waveguides are mathematically considered to be connected to a single point [63]. In reality, however, a single connection point is unphysical in our case as we make use of a realistic VPC junction. This means that some small deviation to the ideal performance can be expected.

To further evaluate this latter scenario of equal signals being directed towards $P_3$ and $P_5$, the input vector $X_1$ was calculated and evaluated numerically. This is done by defining the desired output vector $Y_1 = [0,0,1e^{i\pi},0,1,0]$. Then, the required input vector was calculated at the equal splitting frequency $f = f_S$ directly by inverting $M_S$ so that $X_1 = M_S^{-1}Y_1 \approx [0,1.00,0,0,0,1.00e^{i(3.14)}]$. The performance of the junction excited with the input vector $X_1$ was then evaluated numerically using two point dipoles placed at $P_2$ and $P_6$ (see methods for more details). With this configuration, the numerical result of the real component of the $H_z$ field distribution at $f = f_S$ is shown in Figure 5b(i), demonstrating that the incident signals applied from $P_2$ and $P_6$ are redirected towards $P_3$ and $P_5$, as expected. For completeness, zoom-in snapshots of the $H_z$ field distributions at each port are presented as inset in Figure 5b(i). The red(blue) arrows outside the inset represent the input(output) signals. The black arrows within the inset have been positioned along each port and are rotated so that they point perpendicular to each VPC-VPC waveguide from VPC2 towards VPC1, this is to aid in the comparison of phase between ports. As observed, the two inputs ($P_2$ and $P_6$) and two outputs ($P_3$ and $P_5$) are out-of-phase, as expected from the designed $Y_1$. Finally, the analytical (solid lines) and numerical (symbol) results of $|y_{P_m}|^2$ and $\phi_{P_m}$ are presented in Figure 5b(ii-iii) for a $\pm 2.5$ THz frequency range centered at $f_s$. As expected, there is zero transmission towards $P_{2,4,6}$ and a good agreement for the two outputs towards $P_3$ and $P_5$.

A second example consists of a *wave director* defined by the output vector $Y_2 = [0,0,0,1,0,0]$, meaning that now all the signals are to be directed towards a single port (in this case $P_4$ as an example). Following the same procedure as above, the required input vector is $X_2 \approx [0.58, 0, 0.58 e^{i(2.10)}, 0, 0.58 e^{i(2.10)}, 0]$. The results are shown in Figure 5b(iv) showing how the incident signals are *routed* towards $P_4$ only. Additionally, as it can be seen in Figure 5b(v-vi), both $|y_{P_m}|^2$ and $\phi_{P_m}$ show good agreement between analytic and numerical results. With the examples in this section, it was shown that the retrieved frequency dependent scattering matrix, $M_S$, can be used to analytically model the splitting of incident signals at the 6-port topological junction. These results could be used to design large photonic circuits, for instance, for linear computing processes, avoiding the necessity of large and computationally expensive numerical-based trial and error design processes.

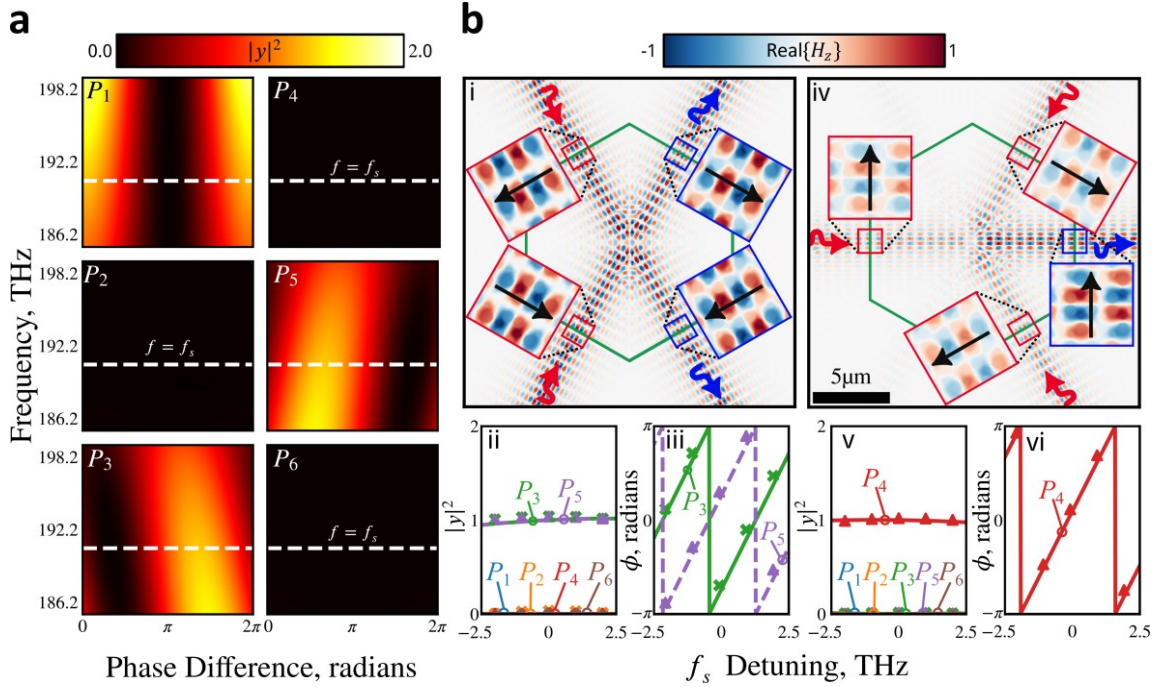

**Figure 5.** Multiple input study. (a) Analytically calculated output magnitude, $|y_{P_m}|^2$, for $m = 1, \ldots, 6$ calculated with an input vector of $X$ for input phase differences $0 \leq \Delta\phi \leq 2\pi$. (b) i-iii) Two input case with designed input vector $X_1$, with results presented in panel (i) of numerically retrieved Re{$H_z$} field distribution at $f = f_s$ with red(blue) arrows representing inputs(outputs) and inset panels displaying the local fields at the input and output ports, with black arrows pointing from VPC2 to VPC1 as spatial reference. (ii) Analytical(lines) and numerical(crosses) results for the input $X_1$ showing retrieved $|y_{P_m}|^2$ and (iii) $\phi_{P_m}$ relative to the input at $P_2$. iv-vi) The same analysis as (i-iii) but for a *wave director* studied with the input vector $X_2$ and $\phi_{P_m}$ relative to the input at $P_1$.

## 5. Conclusion

In this work a 6-port VPC-VPC waveguide junction has been proposed and studied, showing its potential for linear computing operations. An in-depth study has been presented including the analysis of the bulk band structure of single VPCs as well as the design and evaluation of the VPC-VPC waveguide supercell band structure. A frequency dependent scattering matrix of the structure (6-port VPC-VPC waveguide) was then extracted within the spectral range where unique edge state wavevectors exist. Due to the linearity of the designed structure, it has been demonstrated how the output signals can be controlled and arbitrary engineered by simply exploiting the extracted scattering matrix and calculating, fully analytically, the required input signals to achieve any desired output vector of signals. We envisage that our work may be exploited to design larger junction networks for other computing operations including solution of partial differential equations and routing of information.

## 6. Methods

### 6.1 Band Structure

For the bulk VPC, as studied in section 2.1, the TE photonic bands were retrieved using the open-source Python package, MiT Photonic Bands [81], [82]. The two-dimensional model was designed with square mesh cells of uniform size. The side length of all cells, $\Delta d$, was defined relative to the smallest hole radius, so that $\Delta d = (2r_1)/20$ (20 mesh cells per smallest hole diameter). Each VPC unit cell was designed with periodic boundary conditions on all sides (a schematic of 4-unit cells with lattice vectors labeled is shown in the left panel of Figure 1a). With this configuration the first four bands were retrieved for 120 interpolated wavevectors spanning the high symmetric points $\Gamma \to K \to M \to \Gamma \to K' \to M \to \Gamma$.

For the supercell band diagrams in section 2.2, the eigenfrequency solver of the optics module in COMSOL Multiphysics was used to retrieve the TE bands by designing the supercell in Figure 3b(i) (dimensions listed in section 2.2). The physics-controlled mesh option was selected with "Normal" element size preset to construct a triangular mesh based on the frequency independent permittivity of $\varepsilon_{Si}$ and $\varepsilon_{Air}$. To simplify the model (as it is symmetric across the $x$-axis) only the top half of each supercell was designed. The boundary conditions of each supercell included periodic boundary conditions on both the left and right ($\pm x$) boundaries, scattering boundary conditions on the upper ($y+$) boundary and a PMC symmetry plane on the lower ($-y$) boundary. With this configuration the eigenfrequencies were retrieved within the region $155 \leq f \leq 255$ for 100 equally spaced wavevectors between $-0.5 \leq k_x \leq 0.5$.

### 6.2 6-Port Magnitude and Phase Retrieval

MEEP [81] was implemented for the numerical analysis with a uniform mesh cell size of $\Delta d$ as above. Both the 6-port and the straight VPC-VPC waveguide were designed with a square simulation area of side length $d_{sim}$ as well as an additional absorbing layer of length $43a$ on all sides [83] before terminating it with a perfectly electric conductor boundary condition. Point dipole sources were used to excite the edge states. They were placed at the VPC-VPC interfaces and located $30.5a$ from the center of the simulation area (always at the mid-point between four large holes along the VPC-VPC interface). The point sources were orientated so that the electric field component was parallel to the VPC-VPC interface (i.e. for $P_1$ the $E_x$ component was excited). For models with a single input at $P_1$, mirror symmetry along the $y=0$ line with $+1$ phase factor was used, for multiple input examples no symmetry was used. The sources were configured with a central frequency of $\approx 192$ THz and bandwidth of $\approx 25$ THz. Discrete Fourier transform (DFT) flux monitors were placed along each port with length $\frac{2}{\sqrt{3}}d_{hex}$ at a distance of $d_{hex}$ from the junction center (forming the green hexagon in Figure 4b) and monitor 200 equally spaced frequencies in the bandwidth of the source. In addition, a DFT area monitor was placed across the whole simulation domain and monitored 31 equally spaced frequencies, this data was used for phase retrieval and $|H_z|$ plots. The simulation was run until the fraction of energy left in the system compared to its maximum was $1 \times 10^{-6}$.

The scattering parameters were retrieved by integrating the magnitude of the time averaged Poynting vector along each of the DFT flux monitors for each port. Then the spectra were normalized with respect to a straight VPC-VPC waveguide with the same dimensions and a DFT flux monitor at equivalent position to $P_4$ of the 6-port model. For reflected signals at an input port, the input signal was removed using the input vector after normalization, i.e. $\left|y_{P_m}^{\text{reflected}}\right|^2 = \left|y_{P_m}\right|^2 - \left|x_{P_m}\right|^2$; where, $\left|y_{P_m}\right|^2$ is the total normalized scattering parameter including the input and any output signals at the $P_m$ port (in the case of a single input at $P_1$, then $\left|x_{P_1}\right|^2 = 1$). The phase, $\phi$, for each port was retrieved from the area DFT monitors and extracted from $H_z$ along the same port lines as the DFT flux monitors and normalized with the phase measured at $P_1$ of the straight waveguide. The phase was then averaged for a small cross section of each port, corresponding to the $0.49a \leq x_{\text{port}} \leq 0.51a$, where $x_{\text{port}}$ corresponds to the relative position along each port. Here, $x_{\text{port}} = 0$ corresponds to the center of each port and $+x_{\text{port}}(-x_{\text{port}})$ corresponds to the right(left) hand side of a port when facing the direction of propagation (red/blue arrows in Figure 4 and Figure 5). This method was chosen to sample the phase at each waveguide at a position where the $|H_z|$ was near its maximum and to avoid the many nodes present in the edge state that may induce numerical errors in the phase (see Figure 3b for mode profile).

## 7. Supporting Information

Supporting Information is available from the Wiley Online Library or from the author. 1) 6-port structure with a type II VPC-VPC interface. 2) Magnetic field distributions at additional frequencies. 3) The $d_{\text{hex}}$ dependence on the scattering parameters. 4) Effect of removing holes at the junction.

**Acknowledgements**

This work was supported by the Leverhulme Trust under the Leverhulme Trust Research Project Grant scheme (No. RPG-2020-316 and RPG-2023-024). For the purpose of Open Access, the authors have applied a CC BY public copyright license to any Author Accepted Manuscript (AAM) version arising from this submission.

**Conflicts of interests**

The authors declare no conflicts of interests.

**Data availability**

The datasets generated and analysed during the current study are available from the corresponding author upon reasonable request.